
\input harvmac
\input epsf
\noblackbox
\def\L{\Lambda}
\def\a{\alpha}
\def\b{\beta}
\def\ta{\theta_\a}
\def\tb{\theta_\b}

\def\zt#1{{z_#1}}
\def\tu{{\tilde u}}
\def\cL{{\cal L}}
\def\br{\hfill\break}
\def\frac#1#2{\relax{\textstyle {#1 \over #2}}\displaystyle}

\lref\stdual{A. Font, L. Ibanez, D. L\"ust, and F. Quevedo,
     {\it Phys. Lett.} {\bf B249} (1990) 35;
J. Schwarz, hep-th/9307121; \br
A. Sen, {\it Int. J. Mod. Phys.} {\bf A9} (1994) 3007; \br
C. Hull and P. Townsend, Nucl.\ Phys.\ {\bf B438} (1995) 109,
  hep-th/9410167; \br
M. Duff, {\it Nucl. Phys.} {\bf B442} (1995) 47, hep-th/9501030;\br
P. Townsend, {\it Phys. Lett.} {\bf B350} (1995) 184,
hep-th/9501068;\br
E. Witten, Nucl.\ Phys.\ {\bf B443} (1995) 85, hep-th/9503124;\br
A. Sen, hep-th/9504027;\br
J. Harvey and A. Strominger, hep-th/9504047;\br
A. Strominger, hep-th/9504090;\br
B. Greene, D. Morrison and A. Strominger, hep-th/9504145;\br
C. Vafa, hep-th/9505023; \br
C. Vafa and E. Witten, Nucl.Phys. {\bf B447} (1995) 261,
hep-th/9505053;\br
C. Hull and P. Townsend, hep-th/9505073;\br
S. Kachru and C. Vafa, hep-th/9505105;\br
S. Ferrara, J. Harvey, A. Strominger, and C. Vafa, hep-th/9505162;\br
D. Ghoshal and C. Vafa, hep-th/9506122; \br
G. Papadopoulous and P. Townsend, hep-th/9506150;\br
A. Dabholkar, hep-th/9506160;\br
C. Hull, hep-th/9506194;\br
J. Schwarz and A. Sen, hep-th/9507027;\br
C. Vafa and E. Witten, hep-th/9507050;\br
E. Witten, hep-th/9507121;\br
K.\&M. Becker and A. Strominger, hep-th/9507158;\br
J. Harvey, D. Lowe, and A. Strominger, hep-th/9507168;\br
D. Jatkar and B. Peeters, hep-th/9508044; \br
A. Sen and C. Vafa, hep-th/9508064;\br
T. Banks and M. Dine, hep-th/9508071; \br
S. Kachru and E. Silverstein, hep-th/9508096;\br
J. Schwarz, hep-th/9508143;\br
S. Chaudhuri and D. Lowe, hep-th/9508144.
}
\lref\Ntwodual{
A. Strominger, hep-th/9504090;\br
B. Greene, D. Morrison and A. Strominger, hep-th/9504145;\br
C. Vafa and E. Witten, Nucl.Phys. {\bf B447} (1995) 261,
hep-th/9505053;\br
S. Kachru and C. Vafa, hep-th/9505105;\br
S. Ferrara, J. Harvey, A. Strominger, and C. Vafa, hep-th/9505162;\br
C. Vafa and E. Witten, hep-th/9507050;\br
S. Kachru and E. Silverstein, hep-th/9508096.
}
\lref\sw{N. Seiberg and E. Witten, {\it Nucl. Phys.} {\bf B426}
(1994) 19,
     hep-th/9407087.}
\lref\sei{See eg., N. Seiberg, hep-th/9506077;\br
    K. Intriligator and N. Seiberg, hep-th/9506084, and references
    therein.}
\lref\KLTY{A. Klemm, W. Lerche, S. Theisen, and S. Yankielowicz,
       {\it Phys. Lett.} {\bf B344} (1995) 169, hep-th/9411048.}
\lref\AF{P. Argyres and A. Faraggi, {Phys. Rev. Lett.}
     {\bf 74} (1995) 3931, hep-th/9411057.}
\lref\ADS{
M. Douglas and S. Shenker, Nucl.\ Phys.\ {\bf B447} (1995) 271,
      hep-th/9503163;\br
   P. Argyres and M. Douglas, hep-th/9505062.}
\lref\KLT{A.\ Klemm, W.\ Lerche and S.\ Theisen, hep-th/9505150.}
\lref\dWKLL{B.\ de Wit, V.\ Kaplunovsky, J.\ Louis and D.\
       L\"{u}st, hep-th/9504006; \br
      I. Antoniadis, S. Ferrara, E. Gava, K. Narain, and T. Taylor,
       hep-th/9504034.}
\lref\CLM{G.-L. Cardoso, D.\ L\"ust and T.\ Mohaupt, hep-th/9507113.}
\lref\CLMold{G.-L. Cardoso, D.\ L\"ust and T.\ Mohaupt,
hep-th/9412209.}
\lref\GoLo{C.\ G\'omez and E.\ L\'opez, hep-th/9506024.}
\lref\relre{A. Ceresole, R. D'Auria, and S. Ferrara, {\it
         Phys. Lett.} {\bf B339} (1994) 71, hep-th/9408036.}
\lref\dual{A.\ Ceresole, R.\ D'Auria, S.\ Ferrara and A.\ Van
Proeyen,
        Nucl\ Phys.\  {\bf B444} (1995) 92,   hep-th/9502072.}
\lref\fer{M. Billo, A. Ceresole, R. D'Auria,
    S. Ferrara, P. Fre, T. Regge, P. Soriani, and A. Van
    Proeyen, hep-th/9506075.}
\lref\KLM{A. Klemm, W. Lerche and P. Mayr, hep-th/9506112.}
\lref\check{
V. Kaplunovsky, J. Louis, and S. Theisen, hep-th/9506110; \br
I. Antoniadis, E. Gava, K. Narain, and T. Taylor, hep-th/9507115.}
\lref\LY{B.\ Lian and S.T.\ Yau,  hep-th/9507151 and hep-th/9507153.}
\lref\CDGP{P. Candelas, X. de la Ossa, P. S. Green and L. Parkes,
{\it Nucl. Phys.} {\bf B359} (1991) 21}
\lref\HKTYII{S.
Hosono, A. Klemm, S. Theisen and S. T. Yau, {\it Nucl. Phys.} {\bf
B433} (1995) 501, hep-th/9406055}
\Title{\vbox{
\hbox{CERN-TH/95-231}
\hbox{HUTP-95/A032}
\hbox{\tt hep-th/9508155}
}}
{Nonperturbative Results on the Point Particle Limit of}
\vskip-1cm
\centerline{{\titlefont N=2 Heterotic String Compactifications}}
\bigskip\bigskip
\centerline{Shamit Kachru$^{a}$, Albrecht Klemm$^{b}$, Wolfgang
Lerche$^{b}$,}
\centerline{Peter Mayr$^{b}$
and Cumrun Vafa$^{a}$}
\bigskip
\bigskip\centerline{\it $^{a}$Lyman Laboratory of Physics, Harvard
University,
Cambridge, MA 02138}
\bigskip\centerline{\it $^{b}$Theory Division, CERN, 1211 Geneva 23,
Switzerland}
\vskip .3in

Using heterotic/type II string duality, we obtain exact
nonperturbative results for the point particle limit
($\alpha^{\prime}
\rightarrow 0$) of some particular four dimensional, $N=2$
supersymmetric compactifications of heterotic strings. This allows us
to recover recent exact nonperturbative results on $N=2$ gauge theory
directly from tree-level type II string theory, which provides a
highly
non-trivial, quantitative check on the proposed string duality. We
also investigate to what extent the relevant singular limits of
Calabi-Yau manifolds are related to the Riemann surfaces that
underlie rigid $N=2$ gauge theory.

\Date{\vbox{\hbox{CERN-TH/95-231}\hbox{\it {August 1995}}}}

 Recently there have been exciting developments in understanding
non-perturbative aspects of string theory through conjectured string
dualities \stdual. In particular, the geometry of moduli spaces of
$N=1,2$ and $4$ supersymmetric string vacua is getting better
understood. Since for $N=4$ the geometry of the moduli space is
uncorrected even non-perturbatively, the $N=1,2$ cases are much more
interesting, as far as shedding light on non-perturbative dynamics of
string theory is concerned. This is also mirrored in the interesting
dynamics of the $N=1,2$ field theories
\refs{\sw{,}\sei{,}\KLTY{,}\AF,\ADS}. In this paper we
show how some of the exact results on the quantum moduli space of
certain $N=2$ string vacua \ref\kv{S. Kachru and C. Vafa,
hep-th/9505105.}\ can reproduce in the point particle limit (where
$M_{planck}\rightarrow \infty$) the exact field theory results of
\sw.\foot {The question of going to the point particle limit has been
addressed before in~\refs{\relre{,}\dual{,}\GoLo{,}\fer{,}\CLM}.}
This provides a truly nonperturbative, quantitative check on the
proposed heterotic/type II string duality.\foot{ It also
substantiates the conjectures that the local analogs of the
Seiberg-Witten Riemann surfaces are given by Calabi-Yau manifolds
\refs{\relre{,}\dual}\ and that space-time Yang-Mills instanton
effects can be described in terms of world-sheet instantons of type
II string theory \KLTY.}

\lref\Next{Work in progress.}
We will concentrate on the two main models studied in \kv, for which
there have already been many non-trivial checks in perturbation
theory \refs{\KLM,\check,\LY}. We will first study in some
detail the rank three model of \kv, which we will call model A, and
then discuss how our results generalize to
the second main model of \kv\ (the rank four model, which we will
call model B). More details, especially concerning the string and
gravitational contributions to the exact nonperturbative effective
action, will appear in a subsequent paper \Next.

\newsec{Description of Model A}

Model A has two equivalent descriptions: We can view it as the
$E_8\times E_8$ heterotic string compactified on $K3\times T^2$,
where we choose the $T$ and $U$ moduli of the two-torus to be equal
so that there is an extra $SU(2)$ gauge symmetry.
We also choose the second Chern class of the $E_8\times E_8\times
SU(2)$ gauge bundle to be $(10,10,4)$, giving a total of $24$ that
equals the second Chern class of $K3$; this is required for
world-sheet anomaly cancellations. This model has $129$
hypermultiplets whose scalars characterize the geometry of $K3$ with
the corresponding bundles on it, and $2$ vector multiplets whose
scalars give the modulus $T$ of the two-torus and the dilaton/axion
field $S$. The dual description of this model is given by a
type IIB (or type IIA) string compactification on a Calabi-Yau
manifold $M$ (or its mirror), with defining polynomial
\eqn\cyhyp{p = {z_1}^{12} + {z_2}^{12} +
 {z_3}^{6} + {z_4}^{6} + {z_5}^{2}
- 12\psi \,z_{1}z_{2}z_{3}z_{4}z_{5} - 2\phi\, {z_1}^6{z_2}^{6}\ ,}
where $z_{i}$ are coordinates of $W\!P^{12}_{1,1,2,2,6}$ and where we
mod
out by all phase symmetries that preserve the holomorphic three-form.
This Calabi-Yau manifold has $h_{11}=128$ and $h_{21}=2$, giving rise
to $2$ vector multiplets (whose scalar expectation values correspond
to $\phi$ and $\psi$ above) and 129 hypermultiplets (including the
type II string dilaton). If we wish to study the moduli space of
vector multiplets, tree-level type II string theory is exact, whereas
if we wish to study the moduli space of hypermultiplets, the tree
level
of the heterotic side is exact. In this paper we consider the moduli
space of vector multiplets, and so we study the classical moduli
space of the type II side spanned by $\phi$ and $\psi$ in the above
defining equation.

\lref\Candelas{P. Candelas, X. De la Ossa, A. Font, S. Katz, and D.
Morrison, {\it Nucl. Phys.} {\bf B416} (1994) 481, hep-th/9308083.}
\lref\Yau{S. Hosono, A. Klemm, S. Theisen, and S.T. Yau, {\it Comm.
Math. Phys.} {\bf 167} (1995) 301.} The classical moduli space of
this model has been studied in great detail in
\refs{\Candelas{,}\Yau}. It is convenient to introduce the variables
\eqn\xy{
x = -{1\over 864} {\phi\over{\psi^{6}}}\ ,\qquad\
y = {1\over \phi^{2}} ~.}
According to the identification of \kv, the $T$ and $S$ fields
of the heterotic side should be identified (in the large
$S$/weak coupling regime) with the special coordinates corresponding
to $x$ and $y$, respectively.  In particular, for large $S$ one has:
\eqn\hetv{ x={1728\over j(T)}+\dots\ ,\ \ \ \ y={\exp}(-S)+\dots}
This identification was in part motivated by the fact that
at $T=i$  the perturbative heterotic model develops
an $SU(2)$ gauge symmetry.  The existence of the $SU(2)$ gauge
symmetry
of heterotic strings
is reflected by the existence of the conifold locus of $M$, which is
given by
\eqn\disc{\Delta\ =\ (1- x)^{2} -  x^{2}y\ =\ 0~.}
For weak coupling, $y\rightarrow 0$, there is a double singularity at
$x=1$ (corresponding to $T=i$). Moreover, for finite coupling
corresponding to finite $y$, there are two singular loci for $x$, in
line with the field theory results of \sw\ where one has two singular
points in the moduli space associated with massless monopoles/dyons.

\newsec{What to Expect when Gravity is Turned Off?}

It would be a very non-trivial test of all these ideas if we could
show that in the limit of turning off gravitational/stringy effects,
we would reproduce the results of \sw, where the quantum moduli space
of pure
$N=2$ Yang-Mills theory with $SU(2)$ gauge group has been studied.
This corresponds to considering the point particle limit of strings
obtained by taking $\alpha'\rightarrow 0$. To this end note that the
variable $u=tr \phi^2$ that vanishes at the $SU(2)$ point should, to
leading order, be identified with $x-1$. To make this dimensionally
correct, we must have
\eqn\deu{x=1+\alpha ' u +O(\alpha ')^2 ~.}
Note that as $\alpha'\rightarrow 0$, the full $u$-plane is mapped to
an infinitesimal neighborhood of $T=i$. This in particular means that
the effect of the modular geometry of $T$ is being turned off in this
limit, as one expects. Furthermore, in order for the scale $\Lambda$
of the $SU(2)$ theory to satisfy $\Lambda \ll M_{planck}\sim 1/\sqrt
{\alpha'}$, we should tune the string coupling constant (which is
defined naturally at the string scale) to be infinitesimally small.
Taking into account the running of the $SU(2)$ gauge coupling
constant, we should take, to leading order in $\alpha'\rightarrow 0$:
\eqn\deco{y={\exp}(-S)={\alpha '}^2\Lambda^4 {\exp}(-\hat S)\
\equiv:\ \epsilon^2.}
Thus, by dimensional transmutation the coupling constant of $SU(2)$,
$e^{-\hat S}$, can be traded with the scale $\Lambda$, at which the
$SU(2)$ gauge theory becomes strongly coupled. Note that the conifold
locus \disc\ in the limit $\alpha'\rightarrow 0$ goes to
\eqn\uloc{u^2=\Lambda^4 \,{\exp}(-\hat S)\ ,}
which is the expected behaviour.


Let us recall that $N=2$ supergravity moduli are characterized by a
prepotential $F$, which in our case is a function of $T,S$. Using the
axionic shift symmetry, it is easy to see that it has an expansion of
the form \relre
\eqn\expa{F=ST^2+\sum_{n=0}^{\infty} f_n(T){\rm exp}(-nS)\ ,}
where $f_0$ corresponds to one-loop string corrections and
where $f_n(T) {\rm exp}(-nS)$ is the contribution from the $n$-th
stringy instanton sector. This expansion is most convenient when we
are dealing with large $T$. Since we are interested in $T$ near $i$,
it
is more convenient to shift to $\tilde T\sim(T-i)/(T+i)$
\ref\bcov{M. Bershadsky, S. Cecotti, H. Ooguri, and C. Vafa, {\it
Comm. Math. Phys.} {\bf 165} (1994) 311.}, and consider another
expansion of $F$ given by
\eqn\rexpa{F=S{\tilde T}^2+\sum_{n=0}^{\infty} g_n(\tilde T){\rm
exp}(-nS)
+Q(S,\tilde T)}
where $Q(S,\tilde T)$ is some polynomial of first degree in $S$, and
where we have chosen $g_0\sim {\tilde T}^2 {\rm log}{\tilde
T}+O({\tilde T}^3)$. We now consider turning off gravity by
taking the limit $\alpha'\rightarrow 0$. Note that since both
$\tilde T$ and the variable $a$ defined in \sw\ are good special
coordinates and are proportional to leading order, they have to be
identified via%
\eqn\idnt{\tilde T=\sqrt{\alpha'}\, a\ .}
This is consistent\foot{To be precise, in order to recover the
conventional definition of $a$ in relation to $u$, note that there is
a proportionality constant in this equation related to the second
order expansion coefficient of the $j$-function near $T=i$. We can
avoid this by rescaling $u$ in the definition of \deu\ and redefining
$\Lambda$ in such a way that $\tilde u =u/(\Lambda^2 e^{-\hat S/2})$
is invariant. This will have no effect on the equations below.} with
\deu, and also correctly translates the modular transformation $T\to
-1/T$ to the Weyl transformation $a\to -a$
\refs{\dual,\CLMold,\dWKLL,\CLM}.
Now using \deco\ and \idnt\ we reexpand \rexpa\ and get
\eqn\Freex{F=\alpha'{\hat S} a^2+\sum_{n=0}^{\infty}
g_n(\sqrt{\alpha'} a)
{\alpha '}^{2n}\Lambda^{4n}
 {\rm exp}(-n\hat S)+Q({\hat S},\sqrt {\alpha'} a)\ .}
In order to recover the results of Seiberg and Witten
as $\alpha'\rightarrow 0$, we must find for large $a$:
\eqn\Flara{g_n(\sqrt{\alpha'}a)=c_n
(\sqrt{\alpha'} a)^{2-4n}+O((\sqrt{\alpha'}
a)^{2-4n+1})\ ,}
where $c_n$ are the instanton coefficients of \sw\ in the weak
coupling regime. In other words, let $F_{SW}(a,\Lambda^4)$ be the
prepotential obtained in \sw. Then, if $g_n$ behaves as above, we
would have
\eqn\prep{F(a,\hat S)=\alpha'F_{SW}(a,\Lambda^4 {\rm exp} (-\hat S
))+
{\tilde Q}({\hat S},\sqrt{\alpha '} a)+O(\alpha ')^{3/2}\ ,}
where $\tilde Q$ is a quadratic polynomial of first degree in $S$ and
second
degree in $\sqrt{\alpha '}\, a$. In order to compare the above result
with the periods of $M$ that we will determine in this limit below,
it is useful to recover, using special geometry, the periods from the
prepotential $F$ given in \prep. Since we are working in a
non-homogeneous basis, the $A$-type periods can be taken to be
(proportional to)
\eqn\aper{(1,S=\hat S -{\rm log}
(\Lambda^4\alpha'^2),\sqrt {\alpha'} a)\ ,}
while the corresponding $B$-type periods are given by
\eqn\bper{(2F-S\partial_S F-a\partial_aF,\partial_S F,{1\over
\sqrt{\alpha '}}\partial_a F)\ .}
This simplifies due to the remarkable fact (whose full physical
significance remains to be uncovered) proven in \ref\mat{M. Matone,
hep-th/9506102.}\ that
\eqn\mato{\pi i(F_{SW}-{1\over 2}a\partial_a F_{SW})=
(-2\pi i)\partial_{S}F_{SW}=u\ .}
This is crucial for us in obtaining the rigid theory
 in the limit of turning off gravity.
Using \mato\ we
find that the $B$ periods must be certain linear combinations of
the $A$ periods with ${\sqrt \alpha' }a_D,\alpha' u, \alpha' u{\hat
S}$. Thus, up to linear combinations, we should get the following 6
periods
\eqn\alper{(1,S,\sqrt {\alpha'} a,\sqrt{\alpha'}a_D,\alpha'u,
\alpha' u S)\ .}
We will verify below that the periods of the Calabi-Yau manifold $M$
in the limit $\alpha'\rightarrow 0$ are indeed given by linear
combinations of the above six periods.

\newsec{Geometrical Characterization of the Appearance of
$(a,a_{D})$}

 Before going on in the next section to solve the Picard-Fuchs
equations and obtain the six periods that we expect to emerge in the
point particle limit, we would like to give a geometrical idea of how
the two most interesting periods, namely the rigid periods
$a(u),a_D(u)$ of \sw, appear. Given the fact that in \sw\ $a,a_D$
were periods of a meromorphic one-form on a torus, it is important to
see, by a means more transparent than direct computation, how this
geometrical structure is encoded in the Calabi-Yau manifold $M$ in
the vicinity of $y = 0$, $x = 1$.

As we approach the conifold locus in moduli space, some three-cycle
is shrinking to zero size. We expect that the computation of
$(a,a_{D})$ is only affected by integrals localized in the
neighborhood of the collapsing cycle. Therefore, we should try to
understand the appearance of the rigid periods $(a,a_{D})$ by
approaching $y=0$, $x=1$ in moduli space and by simultaneously
rescaling variables to ``blow up'' a neighborhood of the singular
locus on the manifold $M$.

{}From the above we know that as we approach the point of interest,
the moduli of $M$ scale as\foot{In the following, we use the
dimensionless variable $\tu\equiv u/(\L^2 e^{-\hat S/2})$.}
\eqn\scalp{\phi = -{1 \over \epsilon }\ ,\qquad\ \psi =
-{1\over\sqrt3}\big(2^5\epsilon\,(1+\epsilon \tu)\big)^{-1/6}~.}
If we want to keep the conifold singularity at a finite point in
our rescaled variables, fixing $z_{1}$ it turns out that
the unique choice of rescaling is
$\tilde z_{3,4} = \epsilon^{1/6}z_{3,4}$,
$\tilde z_{5} = \epsilon^{1/2}z_{5}$.
Then, if we in addition define
$\zeta = z_{1}z_{2}$,
and rescale $z_i$ by irrelevant numerical factors,
we find that the defining equation $p=0$ of $M$ can be rewritten as
\eqn\leading{p_{0} = {1\over6}\tilde {z_3}^{6} + {1\over6}\tilde
{z_4}^{6} + {1\over6} \zeta^{6}+ {1\over2}\tilde {z_5}^{2} +
\zeta\,\tilde
z_{3} \tilde z_{4}\tilde z_{5} = \epsilon}
%
\eqn\sub{p_{1} = {1\over12}\,{z_1}^{12} + {1\over12}\,{z_2}^{12} +
{\tu\over6} z_{1}z_{2}\tilde z_{3} \tilde z_{4}\tilde z_{5} = -1~.}
Here we have simply used that as $\epsilon \rightarrow 0$, the
defining polynomial can be written as $p =
{1\over \epsilon}p_{0} + p_{1}$.

As $\epsilon$ goes to zero, the leading singularity is described by
$p_0=\epsilon$. However, note that this is itself a
singular space! It is quite clear that the periods that we are
interested in are governed by the subleading piece \sub, which
smooths out the singularity in \leading\ for finite $\epsilon$. More
concretely, we are suggesting that the periods related to
three-cycles that are not collapsing as we approach $y=0, x=1$
should be controlled by the leading term in the $\epsilon$-expansion.
On the other hand, the $\tu$-dependent periods, $a,a_{D}$, are
governed by the sub-leading term, $p_{1}$, which is the first
$\tilde u$-dependent term in the $\epsilon$-expansion. Therefore, in
order
to study the periods $a,a_{D}$, it should be enough to focus on the
variation of $p_{1}$ with $\tu$. Furthermore, since we are interested
in the leading behavior in the $\epsilon$ expansion, we can solve for
$\tilde z_{3,4,5}$ in \leading\ at the singularity.  This will
be more fully justified below.

Thus solving for the singular locus in $p_{0}=0$, we  find
$\tilde z_{3,4} = z_{1}z_{2}$,
$\tilde z_{5} = -(z_{1}z_{2})^{3}$,
and substituting this into $p_1$, we see that the manifold whose
Hodge
variation must give $(a,a_{D})$ is the curve
%
\eqn\family{{1\over12}\,{z_1}^{12} + {1\over12}\, {z_2}^{12} -
{1\over 6}\tu\,
z_{1}^{6}z_{2}^{6} ~ = ~1~.}
We notice that  this curve is very similar to the following
$\Gamma(2)$ torus in $W\!P^{3}_{1,1,2}$ (in the
patch where $w_{3} = 1$),
%
\eqn\gammatwo{{1\over4}{w_1}^{4} + {1\over4}{w_2}^{4} -
{1\over2}{\tu}\,
w_{1}^{2}w_{2}^{2} ~=~{w_3}^{2}\ ,}
which underlies the rigid $SU(2)$ $N=2$ gauge theory \sw. Note, in
particular, that the two curves share the same discriminant locus,
$\tu=\pm1$. One may in fact view \family\ as a triple cover of the
Seiberg-Witten torus.\foot
{
A related point was made in \fer\ where it is observed that a triple
cover of
\gammatwo\ (for $\tilde u=0$) appears as the locus $z_{3}=z_{4}=0$ in
$M$.
}
\lref\arn{V.I. Arnold, S.M. Gusein-Zade, and A.N. Varcenko, {\it
Singularities of Differentiable Maps}, Vol. I (Birkh\"auser, Boston,
1985); \br V.I. Arnold, S.M. Gusein-Zade, and A.N. Varcenko, {\it
Singularities of Differentiable Maps}, Vol. II (Birkh\"auser, Boston,
1988).} \lref\cec{S. Cecotti, {\it Nucl. Phys.} {\bf B355} (1991)
755; S. Cecotti, {\it Int. J. Mod. Phys.} {\bf A6} (1991) 1749.}

Obtaining the geometrical torus, or more precisely a geometrical
object with equivalent Hodge variation, is not quite enough to yield
the periods $a$ and $a_D$. Another ingredient in \sw\ was the choice
of a particular {\it meromorphic} one-form $\lambda$, whose periods
are $a$ and $a_D$. The meromorphic one-form has no residue and
satisfies $\partial_u \lambda =\omega$, where $\omega$ is the
holomorphic one-form on the torus. How does this emerge for us? In
order to address that and make our discussion of the limit $\epsilon
\rightarrow 0$ above somewhat more rigorous, we use the definition of
periods given in \refs{\arn{,}\cec}. That is, we write the periods
of $M$ as
\eqn\Period{\Pi_i=\int_{\gamma_i}dz_1\,dz_2\,d{\tilde z}_3\,d{
\tilde z}_4\,d{\tilde z}_5 ~{\rm exp}(iW)\ ,}
where $W$ is given by the defining polynomial in weighted projective
space, and $\gamma_i$ are a basis for an appropriate class of cycles
\arn. In our case, $W={p_0\over \epsilon}+p_1$.  We can now
reformulate
what
we were doing before: In the limit as $\epsilon \rightarrow 0$ the
leading contribution to the integral comes from going to the saddle
points of $p_0$ (this is of course nothing but the stationary phase
method). Thus we have to find the `minima of the action', which means
solving
$$\partial_i p_0=0 \ .$$
The minima of the action are not isolated, because
$p_0=0$ gives a singular manifold. Shifting
$$\hat z_{3,4}=\tilde z_{3,4}-\zeta \ ,\ \  \ \ \hat z_5=
\tilde z_5+\zeta^3\ ,$$
we find that
$${1\over \epsilon}p_0={1\over \epsilon}\big[{15\over 6}\zeta^4(\hat
z_3^2+\hat
z_4^2) - \zeta^{4} \hat z_{3}\hat z_{4} + {1\over 2}\hat z_5^2 +
\zeta^{2}(\hat z_{3}\hat z_{5} + \hat z_{4}\hat z_{5}) + O(\hat
z_i^3)\big]$$
It is easy to see that for fixed $\zeta \neq 0$ the `mass matrix' for
$\hat z_{3,4,5}$ has nonvanishing determinant. Thus the `fields'
$\hat z_{3,4},\hat z_5$ for generic $\zeta$ are infinitely
massive\foot{This is true for all $\zeta$ except $\zeta=0$ -- it may
be that
the other four periods are related to this contribution.} as
$\epsilon \rightarrow 0$. We can therefore integrate them out, and
this results, in leading order, in substituting $\hat z_{3,4}=\hat
z_5=0$ in $p_1$, which is precisely what we did above. From the
gaussian integration we get in addition a factor of $\zeta^{-4}$, so
that
$$\Pi =\int (z_1z_2)^{-4}\,dz_1\,dz_2~{\rm exp}(i \big[ {1\over 12}
z_1^{12}
+{1\over 12}z_2^{12}-{1\over 6}\tu z_1^6z_2^6\big])\ .$$
In order to make contact with geometry, we are at liberty to add an
extra $\tu$-independent integral, since multiplying $\Pi$'s by
overall constants is irrelevant. We thus choose
$$\Pi=\int (z_1z_2)^{-4}~v_3^2~ dz_1\,dz_2\,dv_3 ~{\rm exp}(i \big[
{1\over 12}
z_1^{12} +{1\over 12}z_2^{12}-{1\over 6}\tu z_1^6z_2^6+v_3^2\big])$$
(the choice of the $v_3^2$ term in front of the measure will be
explained
momentarily). In the computations of the periods we are free to make
any birational transformation, so we choose $v_1=z_1^3$, $v_2=z_2^3$,
and obtain (after irrelevant rescaling of variables)
$$\Pi =\int (v_1v_2)^{-2}\,v_3^2~ dv_1\, dv_2 \,dv_3 ~{\rm
exp}(i\big[
v_1^{4}+v_2^4-2\tu v_1^2v_2^2+v_3^2\big])$$
Note that the choice of $v_3^2$ above makes the factor in front scale
invariant, and it is thus a function on the resulting elliptic curve.
We then find that the above period is the same as that of the
following meromorphic one-form (by going to the $v_2=1$ patch and
rewriting $v_1\rightarrow x$, $v_3\rightarrow y$):
\eqn\merolam{
\eqalign{
\lambda\  &=\ {y^2\over x^2}{dx\over y}\cr
y^2 \ &\equiv\ x^4-2\tu x^2+1 \ .}}
Note that this meromorphic one-form has only second order
poles, so its periods make sense, and also that
$${\partial \lambda\over \partial \tu}={dx\over y}$$
which is a necessary requirement \sw. In fact, these two ingredients
fix $\lambda$ up to an exact form.
After the
transformation \ref\bate{{\it Higher Transcendental Functions},
Vol. II, A. Erdelyi, ed., Robert E. Krieger Publishing Co., 1981.}\
which brings the quartic torus \merolam\ and the cubic torus of \sw\
into Weierstrass form one can show that $\lambda$ differs from the
meromorphic one form of \sw\ only by an exact piece. Thus $a,a_D$
computed from the periods of $\lambda$ agree with those of \sw.

\newsec{Specialization of Picard--Fuchs Operators}

In the previous section we have seen how two of the periods of $M$
reproduce the expected point particle limit periods $(a,a_D)$. In
this section we will show, by solving the Picard--Fuchs equations,
that all six periods have indeed the expected form \alper.

Solutions of the relevant Picard-Fuchs equations have been computed
in \Yau, and provide the instanton-corrected period vector in the
large complex structure limit $x=0,\ y=0$ ($T\to
i\infty,S\to\infty$). However, for comparison with the rigid $SU(2)$
Yang-Mills theory, we need to expand the periods around $x=1,\ y=0$
($T=i,S\to\infty$), a point which is outside the radius of
convergence of the solutions given in \Yau. Therefore, we will make a
variable transformation to solve the Picard--Fuchs system directly in
variables centered at $x=1,\ y=0$. This is the point of tangency
between the conifold (monopole) locus, $\Delta=0$, and the
weak-coupling line, $y=0$ (see \Candelas\ for details of the moduli
space).

In order to obtain appropriate solutions in form of ascending power
series, partly multiplied by logarithms, we have to be careful in the
choice of variables. A proper way to do that is to blow up twice the
point of tangency by inserting $P^1$s. This leads to divisors with
only normal crossings, and the associated variables will
automatically lead to solutions of the desired form. More precisely,
as shown in Fig.1, the blow-up introduces two exceptional divisors,
$E_2$ and $E_3$. The latter can be associated with the $SU(2)$
Yang-Mills quantum moduli space, which is given by the $u$ plane. It
intersects with the other divisors at the points
$\tu=\infty,\,\tu=\pm1$ and $\tu=0$, corresponding to the
semi-classical limit, the massless monopole points and the $Z_2$
orbifold point, respectively.

\goodbreak\midinsert
\centerline{\epsfxsize 3.6truein\epsfbox{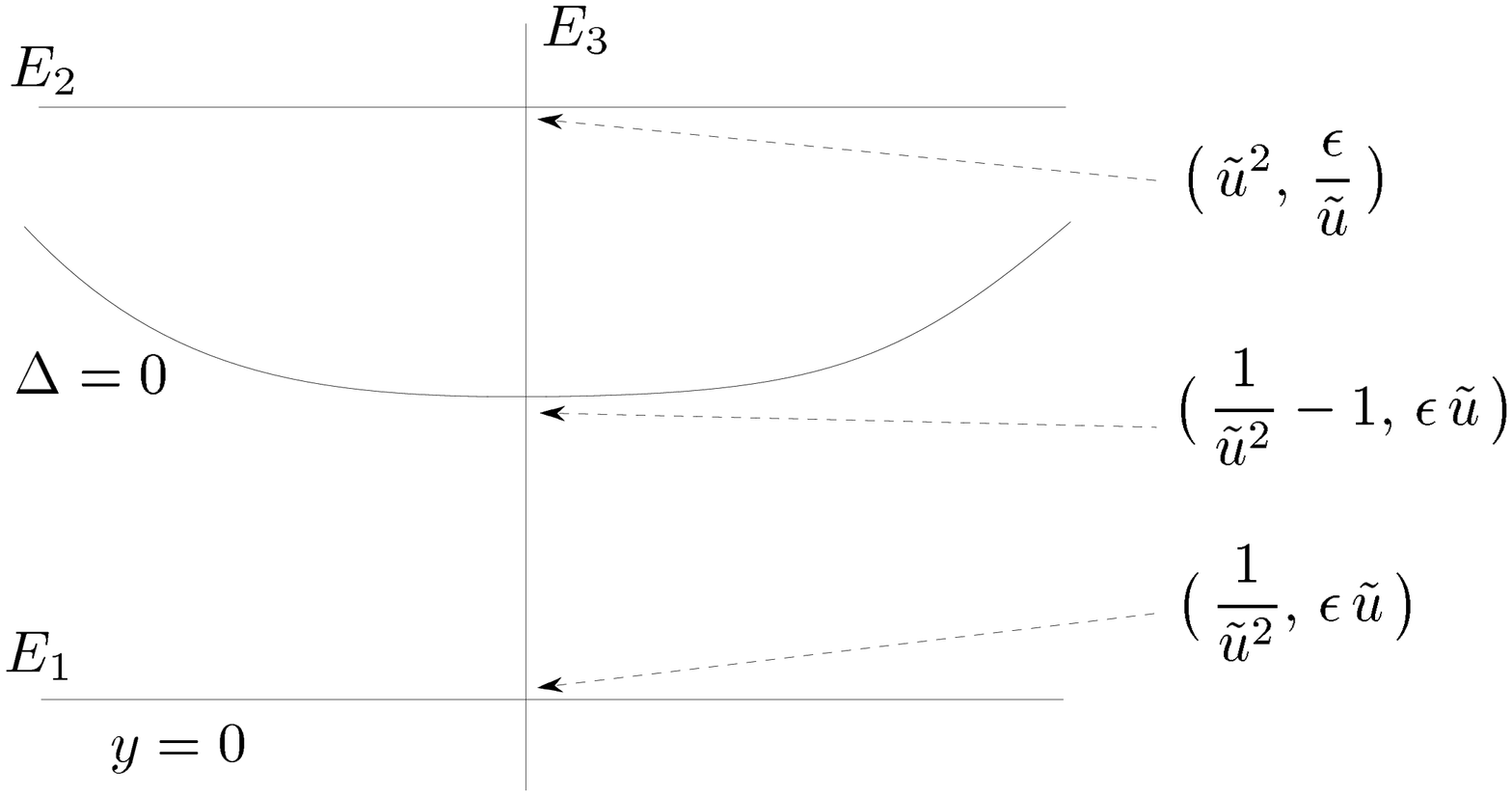}}
\smallskip
\leftskip 2pc \rightskip 2pc
\noindent{
\ninepoint\sl \baselineskip=8pt
{\bf Fig.1} The double blow-up of the intersection of the conifold
locus $\Delta=0$ with $y=0$ leads to three divisor crossings and thus
to three canonical pairs of expansion variables $(x_1,x_2)$. They
describe the physical regimes of the Seiberg-Witten theory at
$\tu=0$, $\tu=\pm1$ and $\tu=\infty$, respectively.}
\smallskip\endinsert

To recover the rigid periods $a(\tu), a_D(\tu)$, we consider for the
time being the specialization of the Picard--Fuchs system only in the
semi-classical regime, $\tu\to \infty$, which corresponds to the
intersection of $E_1$ with $E_3$; the other two regimes can be
treated in a completely analogous way. The appropriate variables are
$x_1=x^2y/(x-1)^2= 1/{\tu}^2$ and $x_2=(x-1)=\a'u\equiv\epsilon\,
\tu$; in particular, $x_2$ is the variable which we will send to zero
when we turn off gravity. After transforming the Picard--Fuchs system
\refs{\Candelas,\Yau}\ to these variables, we find four solutions
with index $(0,0)$ and two solutions with index $(0,1/2)$, where the
index is defined by the lowest powers of the variables (modulo
integers) that appear in the solution. From the monodromy around
$\tu=\infty$ it is clear that the solutions related to the rigid
$SU(2)$ periods must be those with index $(0,1/2)$. Specifically, we
find for the first Picard--Fuchs operator in the limit $x_2 \to 0$
the following leading pieces:
$$
\eqalign{
{\cal L}_1\ \sim\ &
288\,{{ x_1 }}^{2}{ \partial_2 }\,{{ \partial_1 }}^{2}{ x_2 }+288\,{
x_1 }\,{ x_2 }\,{ \partial_2 }\,{ \partial_1 }-144\,{{ x_1 }}^{2}{{
\partial_1 }}^{2}+72\,{{ x_2 }}^{3}{{ \partial_2 }}^{3}-216\,{ x_1 }
\,{ \partial_1 }\cr
&-288\,{ x_1 }{{ x_2 }}^{2}\,{ \partial_1 }\,{{
\partial_2 }}^{2}+108\,{{ x_2 }}^{2}{{ \partial_2 }}^{2}
%
%
\ .}$$
Note that this vanishes identically when applied to a
function of the form $\sqrt{x_2}f(x_1)$. On the other hand, after
rescaling the solutions by ${x_1}^{1/4} {x_2}^{1/2}$ (which is
motivated by the form of the solutions given below), the second PF
operator becomes:
$$
{\cal L}_2\  =\
-1 + 24\,\left(  x_1-1 \right) \,x_1\,\partial_1 +
  16\,\left(  x_1-1 \right) \,{{x_1}^2}\,{{\partial_1}^2} -
  16\,{{x_1}^2}\,x_2\,\partial_1\,\partial_2 +
  4\,x_1\,{{x_2}^2}\,{{\partial_2}^2}\,.
$$
For $x_2\to0$, this becomes precisely the PF operator $\tilde{ {\cal
L} }$ of the rigid $SU(2)$ theory \KLT\ that has $a,a_D$ as solutions
! \foot{ From the form of solutions given below one can see that
there are no logarithms of $x_2$ in the relevant solutions, so that
the $x_2$ dependence in ${\cal L}_2$ cannot cancel out when acting on
them. Note also that $\tilde\cL$ is obtained from the differential
operator ${\cal L}$ that acts on the ordinary torus periods
$(\omega,\omega_D)\equiv (\partial_u a,\partial_u a_D)$ by
$\partial_u \tilde{\cL}= \cL \partial_u$.} Note, in addition, that
for $x_2\to0$, ${\cal L}_2\cdot\lambda=0$ (modulo an exact form),
confirming our choice of meromorphic one-form in \merolam.

Having thus explicitly shown that the Seiberg-Witten periods
$a(u),a_D(u)$ appear as solutions of the Calabi-Yau Picard-Fuchs
system in the limit $x_2\to0$, it remains to verify that the
structure
of the full Calabi--Yau period vector is consistent with our physics
expectations. Indeed, the leading terms of the six solutions in the
limit $x_2 \to 0$ are given by:
\def\al{\alpha}\def\pr{\prime}
\def\ap{\alpha^\prime}

\def\cO{ {\cal O} }

\eqn\PFsol{\eqalign{
1+\cO(x_2^2)&=1+\cO(\al^{\pr 2}u^2)\cr
x_2+O(x_2^2)&=\ap u + \cO(\al^{\pr 2}u^2)\cr
\sqrt{x_2}(1+\cO(x_2))(1-{1\over 16}x_1-{15\over1024}x_1^2+...)
&=\sqrt{\ap}a(\tu^2)(1+\cO(\al^{\pr}u))\cr
(1+\cO(x_2^2))\ln(x_1x_2^2)&=-S(1+\cO(\al^{\pr 2}u^2))\cr
x_2(1+\cO(x_2))\ln(x_1x_2^2)&=-\ap u \ S(1+\cO(\al^{\pr}u))\cr
\sqrt{x_2}(1+\cO(x_2))(1-{1\over
16}x_1-{15\over1024}x_1^2+...)\ln(x_1)&=
\sqrt{\ap}a_D(\tu^2)(1+\cO(\al^{\pr}u))
\ ,}}
in perfect accordance with \alper~! (We intend to present the
precise linear combinations that correspond to the geometric periods
elsewhere \Next.) Note that the appearance of odd powers of
$u$ signals the breaking of the discrete $Z_8$ $R$-symmetry of the
rigid Yang-Mills theory to $Z_4$.  This is due to the string winding
modes, which break this symmetry already in string perturbation
theory.

\newsec{Calabi--Yau Monodromies and the Heterotic Duality Group}

\def\bA{{\bf A}}\def\bB{{\bf B}}\def\bT{{\bf T}}

In the previous section we have solved the Picard--Fuchs equations
near $x=1,y=0$. We would now like to find the monodromies of the
Calabi-Yau manifold, which represent non-perturbative quantum
symmetries from the viewpoint of the heterotic string. In particular,
we will show that certain monodromies
reproduce the monodromies of the quantum $SU(2)$ Yang-Mills theory
and thus underlie the Riemann-Hilbert problem whose solution is given
by the Seiberg-Witten periods, $a(u), a_D(u)$.

The calculation of the monodromy generators is
completely analogous to that of the octic discussed in \Candelas, and
we refer to this paper for details and notation. In summary, the
monodromy group is generated by three elements, denoted by
$\bA$,$\bB$, and $\bT$, which are obtained by loops in the moduli
space around the ${\bf Z}_{12}$ identification singularity $\psi=0$,
the strong coupling singularity $y=1$ and the conifold singularity
$\Delta=0$, respectively.

The period vector in an integral basis is determined by the
holomorphic prepotential $F$ as
$\Pi=
(2 F -\sum_i t_i \partial_{t_i} F,
\partial_{t_1}F ,
\partial_{t_2}F ,
1,
t_1,
t_2)$.
By fixing the integral basis as in
\refs{\CDGP,\Candelas,\HKTYII},
we find in the large complex structure limit
the following prepotential:
\eqn\Prep{
F= -{2\over 3}t_1^3-t_1^2t_2+b_1 t_1 + b_2 t_2 + c \  +\
{\rm inst.\  corr.}\ ,}
where $t_1$ and $t_2$ are inhomogeneous special coordinates,
defined as quotients of the periods $t_i := {\omega_i(x,y)\over
\omega_0(x,y)}$. Here, $\omega_0(x,y)$ is the unique power series
solution in the domain around $(x,y)=0$, while $\omega_{1}(x,y)$,
$\omega_{2}(x,y)$ are the unique solutions of the form
$\omega_0\log(x)+x+\ldots$, $\omega_0\log(y)+y+\ldots$. More
precisely, via mirror symmetry $t_1$ and $t_2$ are the complexified
parameters of the K\"ahler classes $J_1$ and $J_2$ that generate
the K\"ahler cone and that correspond to the divisors in the linear
system of degree
two monomials and degree one monomials. As a consequence, the cubic
part\foot{The constants $b_i,c$ are related to topological invariants
\HKTYII\ by $b_i={1\over 24}\int_M c_2\wedge J_i$ and \CDGP\ $c=i
{\zeta(3)\over (2\pi )^3} \int_M c_3$, i.e. in this case
$b_1={13\over 6}$, $b_2=1$ and $c=-{i \zeta(3)\over 2 \pi^3}63$.} of
$F$ is fixed by the classical intersection numbers and is given by
$-{1\over 6} (\int_M J_i\wedge J_j \wedge J_l) t_i t_j t_l$.

The K\"ahler structure parameters can be related to the heterotic
moduli by $t_1=T,\ t_2 =-{1\over 2\pi i}S$. We now identify the
generators that correspond to the semi-classical heterotic duality
symmetries, namely to T-duality and to the dilaton shift. The shift
generators $\bT_i:\ t_i\to t_i+1$ are immediately determined by the
large complex structure limit to be $\bT_1= (\bA\bT\bA\bT)^{-1},\
\bT_2=(\bA\bT\bB)^{-1}$, whereas a generator respecting the weak
coupling limit and acting as ${\bf S}_1:t_1\to -1/t_1$ on the
semi-classical period vector is given by
$\bT^{-1}\bA^{-1}\bT^{-1}\bA$. A generator that acts on the special
coordinates in a particularly interesting\foot{ As was pointed out in
\KLM, this transformation leads to a symmetry that is quite
mysterious from the point of view of heterotic strings, if we choose
the following, alternative identification: $t_2={\tilde S}-T$. For
this identification, ${\bf U}$ simply acts as an exchange of ${\tilde
S} \leftrightarrow T$ !} way is given by ${\bf
U}=\bA\bT\bB^{-1}\bT^{-1}\bA^{-1}:\ t_1 \to t_1+t_2,\ t_2 \to -t_2$.

An explicit matrix representation of the generators of the duality
group in the large complex structure basis is given by:
{
\eqn\monoA{\eqalign{\bA&=\left(\matrix{
                             -1&  0&  1& -2&  0&  0\cr
                              0&  1&  0&  0&  2&  0\cr
                             -1&  1& -1& -1&  2&  1\cr
                              1&  0&  0&  1&  0&  0\cr
                             -1&  0&  0& -1&  1&   1\cr
                              1&  0&  0&  1&  0&  -1}\right),\
          \bT=\left(\matrix{1&0&0&0&0&0\cr
                           0&1&0&0&0&0\cr
                           0&0&1&0&0&0\cr
                          -1&0&0&1&0&0\cr
                           0&0&0&0&1&0\cr
                           0&0&0&0&0&1}\right),}}
$$         \bT_2=\left(\matrix{
                           1&0&-1&2&0&0\cr
                           0&1&0&0&-2&0\cr
                           0&0&1&0&0&0\cr
                           0&0&0&1&0&0\cr
                           0&0&0&0&1&0\cr
                           0&0&0&1&0&1}\right).$$}
To make contact with the results of \sw, we make a further symplectic
change of basis to the string frame introduced in \dual, which is
characterized by a semi-classical period vector of the form
\def\h{{1\over2}}
\eqn\newBas{
\eqalign{
\Pi_{string}\ &\cong\ (-\h T^2+\h,\ \ -T,\ \ -\h T^2 -\h,\ \
-{2\over3}T^3-T^2S-{13\over6}T+S-2c,\cr
&-2T^2-2TS+{13\over6},\ \ {2\over3}T^3+T^2S+{13\over6}T+S+2c) \ .}}
In this basis, the monodromies $\bT_1,\bT_2$ and ${\bf S}_1$ read
{
\eqn\monoB{\eqalign{
\bT_1&=\pmatrix{{1\over 2}&1&{1\over 2}&0&0&0\cr\noalign{\medskip}-
1&1&1&0&0&0\cr\noalign{\medskip}-{1\over 2}&1&{3\over 2}&0&0&0
\cr\noalign{\medskip}-5&2&5&{1\over 2}&1&-{1\over
2}\cr\noalign{\medskip}-2&4&2&-1
&1&-1\cr\noalign{\medskip}5&-2&-5&{1\over 2}&-1&{3\over 2}},\
\bT_2=
\pmatrix{1&0&0&0&0&0\cr\noalign{\medskip}0&1&0
&0&0&0\cr\noalign{\medskip}0&0&1&0&0&0\cr\noalign{\medskip}2&0&0&1&0
&0\cr\noalign{\medskip}0&2&0&0&1&0\cr
\noalign{\medskip}0&0&-2&0&0&1},}}
$${\bf S}_1=
\pmatrix{1&0&0&0&0&0\cr\noalign{\medskip}0&1&0
&0&0&0\cr\noalign{\medskip}0&0&-1&0&0&0\cr\noalign{\medskip}0&0&0&1&0
&0\cr\noalign{\medskip}0&0&0&0&1&0\cr\noalign{\medskip}0&0&4&0&0&-1}
$$
}
Note that ${\bf S}_1={\bf M}_{\infty}$ contains the $SU(2)$
monodromy \sw\ at $u=\infty$ in an unexpectedly simple way: the
non-trivial $2\times2$ matrix acting on the 3rd and 6th entry of the
period vector is precisely $M_\infty$ of the rigid gauge theory, in a
basis with periods $(\partial_u a,2 \ \partial_u a_D)$. In fact, we
can do better and determine also the strong coupling (monopole)
monodromies from the fact that the monodromy around the conifold
locus is $\bT$. In this way, we find ${\bf M}_1=\bT^{-1}$, ${\bf
M}_{-1}=(\bA^{-1}\bT\bA)^{-1}$ with ${\bf M}_1 {\bf M}_{-1} = {\bf
M}_\infty$, where ${\bf M}_1$ and ${\bf M}_{-1}$ are $6\times6$
matrices with the $SU(2)$ monopole monodromies $M_1$ and $M_{-1}$ in
the $(3,6)$ entries as the only non-trivial elements.

The identification of the semi-classical heterotic string
monodromies, as well as of non-perturbative monodromies of the field
theory limit as part of the Calabi--Yau monodromies, provide a
non-trivial check on the type II-heterotic string duality. More
importantly, we get a prediction for the non-perturbative duality
group of the heterotic theory: it is generated by $\bA,\ \bB$ and
$\bT$, subject to certain relations that are implied by the large
complex structure limit \refs{\Candelas,\HKTYII}, and by the
Van--Kampen relations. Specifically, one can show that
$$
\eqalign{
&[\bB, (\bA\bT)^2 ]=[\bB, \bA^2 \bT]=[\bT, \bB
\bA^2]=[\bT,(\bB\bA)^2]=[\bT,
\bA^{-1}\bB\bA]=[\bA^2,\bT\bB]=0 \cr
&(\bA \bT \bB)+(\bA \bT \bB)^{-1} = 2,\ \ \ \bA^6=-1\ .}
$$
It is easy to see that the PF solutions of the previous section are
compatible with the above monodromies.

\newsec{Discussion of Model B}

We would like to discuss how Model $B$ specializes to the rigid
$SU(3)$, $SU(2)\times SU(2)$ and $SU(2)$ $N=2$ Yang-Mills theories,
respectively, near the appropriate points in moduli space;
we will mainly be
concerned with the $SU(3)$ point, but will also briefly discuss the
two other cases.

\lref\dive{R. Dijkgraaf, E. Verlinde and H. Verlinde, {\it Comm. Math. Phys.}
{\bf 115} (1988) 649.}

On the heterotic side,
Model B is obtained by compactifying the $E_8\times E_8$ heterotic
string on $K3\times T^2$ with the second Chern class of the gauge
bundle chosen to be $(12,12)$. This model has three vector multiplets
containing $S,T,U$, and 244 hypermultiplets. Note that in
perturbation theory we get in this model an $SU(2)$ enhanced gauge
symmetry on the line $T=U$, an $SU(2)\times SU(2)$ gauge symmetry at
the point $T=U=i$, and an $SU(3)$ gauge symmetry at the point
$T=U=\rho\equiv e^{2\pi i/3}$ \refs{\dive,\CLMold,\dWKLL,\CLM}.

In the dual type II string framework \kv,
the defining polynomial of the Calabi-Yau manifold is:
\eqn\MB{p={1\over24}{z_1}^{24} \!+\! {1\over24}{z_2}^{24}
 \!+\! {1\over12}{z_3}^{12} \!+\! {1\over3}{z_4}^3 \!+\!
{1\over2}{z_5}^2 \!-\!
 \psi_0 z_1z_2z_3z_4z_5-{1\over6}\psi_1 (z_1z_2z_3)^6\!-\!
{1\over12}\psi_2(z_1z_2)^{12}}
%
in $W\!P^{24}_{1,1,2,8,12}$.
Suitable variables are: $x = -\psi_0^6/ \psi_1$, $y
= {1/\psi_2^2}$ and $z = -{\psi_2/\psi_1^2}$, in terms of which the
discriminant is
\eqn\disc{
\Delta\ =\ \big(y-1\big)\Big\{ {{\big( 1 - z \big) }^2} - y\,{{z}^2}
\Big\} \,\Big\{ {\big( \big( 1 - x \big)^2 - z \big)^2} -
\,y\,{{z}^2} \Big\}\ \equiv:\ \Delta_y\,\Delta_z\,\Delta_x\ .}
We are mainly interested in the region of moduli space near the zero
of
$\Delta$ at $x=0,\,y=0$ and $z=1$, which describes the point of
enhanced $SU(3)$ symmetry. We first need to find the relationship
between $x,z$ and the variables $u,v$ of rigid $SU(3)$ Yang-Mills
theory, in the semi-classical domain where $y=0$. For this, we can
make use of the following identification of $x,z$ with modular
functions \refs{\KLM,\LY}: \def\TAU{T}\def\RHO{U}
\eqn\Jfunc{\eqalign{
(x)^{-1} &=  864 {j(\TAU)+j(\RHO)-1728\over j(\TAU) j(\RHO)
+\sqrt{j(\TAU)(j(\TAU)-1728)}\sqrt{j(\RHO)(j(\RHO)-1728)}} \cr z &=
{(j(\TAU) j(\RHO)
+\sqrt{j(\TAU)(j(\TAU)-1728)}\sqrt{j(\RHO)(j(\RHO)-1728)})^2\over
{j(\TAU)
j(\RHO)(j(\TAU)+j(\RHO)-1728)^2}}\ ,
}}
where $\TAU,\RHO$ are the flat coordinates associated with $x,z$.
Specifically, we can expand the $j$-functions near the origin as
follows:
\eqn\jUT{\eqalign{
j(\TAU)\ &=\ c\,\Big({\TAU-\rho\over\TAU-\rho^2}\Big)^3\ \equiv\
({\sqrt {\alpha'}}a_\TAU)^3\cr
j(\RHO)\ &=\ c\,\Big({\RHO-\rho\over\RHO-\rho^2}\Big)^3\ \equiv\
({\sqrt {\alpha'}}a_\RHO)^3\ ,
}}
where $c$ is some constant that we
will neglect in the following. The point is that $a_\TAU,\,a_\RHO$
have a simple relationship to the $SU(3)$ Cartan subalgebra
variables, $a_1,\,a_2$ \CLM:
\eqn\aTAU{a_\TAU\ =\ \rho\,a_1+a_2\ ,\qquad\ \ \ a_\RHO\ =\
-(\rho^2\,a_1+a_2)\ .}
This can be seen by comparing the Weyl- and modular
transformation behavior, and noting that $j=0$ ($\TAU,\RHO=\rho$)
corresponds to the fixed point of the modular transformation ${\bf
ST}$. In particular, the combined $Z_3$ transformations ${\bf ST}$
(acting on $T$) and $({\bf ST})^{-1}$ (acting on $U$) induce a
Coxeter transformation on $a_i$, and $\TAU\leftrightarrow\RHO$ yields
a Weyl reflection. The three lines in the CSA on which there is an
unbroken $SU(2)\times U(1)$ are given by
$a_\TAU=(1,\rho,\rho^2)\cdot a_\RHO$ \CLM.

By taking $a_\TAU,\,a_\RHO$ large (but $\alpha'$ sufficiently small
so that ${\sqrt {\alpha'}}a_{\TAU,\RHO}\ll1$), we can now make use of
the
semi-classical relationship between $a_i$ and the Casimirs:
$u={a_1}^2+{a_1}^2-a_1 a_2$, $v=a_1 a_2 (a_1-a_2)$, and thus express
$x,z$ in terms of $u,v$ via the $j$-functions. In this way,
we are led to write
\eqn\xyz{\eqalign{
x\ &=\
2\,(\alpha 'u)^{{3/ 2}}\cr
y\ &=\ 27 (\alpha ')^3 {\Lambda_S}^6 \equiv: \epsilon^2\cr
z\ &=\
1 - (\alpha')^{3/2}\left( 2\,{u^{{3/ 2}}} +
3\,{\sqrt{3}}\,v \right)\ ,}}
where ${\L_S}\equiv \L\,{\rm exp}(-\hat S/6)$. Indeed, in terms of
these variables, the leading piece of the discriminant \disc\ in the
$\alpha'\rightarrow 0$ limit is given precisely by the discriminant
of SU(3) quantum Yang-Mills theory \refs{\KLTY{,}\AF},
\eqn\suthreedisc{\Delta_{SU(3)}\ =\ ( 4\,{u^3} - 27\,{( v-{\L_S}^3
)^2})\,
(4\,{u^3} - 27\,{( v +{\L_S}^3)^2})\ .}

We now introduce the dimensionless quantities
$\tu=u/(27^{1/6}\L_S)^2$,
$\, \tilde v=v/(27^{1/6}\Lambda_S)^3$, and consider the
following variables which correspond to blowing up the singular point
$x=y=0$, $z=1$ in a particular way:
\eqn\zi{
\eqalign{
x_1\ &=\ { y\over  x^2 }\ =\ {1\over 4\tu^3} \cr
x_2\ &=\ {1\over  x}\big({1- x - z}\big)\ =\
\sqrt{{27{\tilde v}^2\over 4\tu^3}}\cr
x_3\ &=\ {1\over2}x\ =\ \epsilon\, \tu^{3/2}\, .}}
In terms of these variables, and after rescaling the solutions by
${x_1}^{1/6}\sqrt x_3$, the Picard-Fuchs operators (given in \Yau)
take the following form:
$$
\eqalign{
{\cal L}_1\ &=\
x_3 + 24x_1\left( 5x_3-4  \right) \partial_1 +
  72{{x_1}^2}\left(  2x_3-1 \right) {{\partial_1}^2} -
  6\left( 5x_2 + 6x_3 + 6x_2x_3 \right) \partial_2 \cr&-
  72x_1x_2\partial_1\partial_2 +
  18\left( 1 + x_2 \right)
   \left( 1 - x_2 - 2x_3 - 2x_2x_3 \right) {{\partial_2}^2}
   + 6x_3\left(  8x_3-1 \right) \partial_3 \cr&+
  72x_1\left( 1 - 2x_3 \right) x_3\partial_1
   \partial_3 + 36x_2x_3\partial_2\partial_3 +
  18{{x_3}^2}\left(  2x_3-1 \right) {{\partial_3}^2}}
$$
\eqn\Lops{
\eqalign{
{\cal L}_2\ &=\
16x_1{{x_3}^2}-1  + 24x_1
   \left(  11x_1{{x_3}^2}-2 \right) \partial_1 +
  36{{x_1}^2}\left(  4x_1{{x_3}^2}-1 \right)
   {{\partial_1}^2} \cr&+ 12x_1x_3
   \left( 1 - 2x_3 - 2x_2x_3 \right) \partial_2 +
  72{{x_1}^2}x_3\left( 1 - 2x_3 - 2x_2x_3 \right)
   \partial_1\partial_2 +\cr& 9x_1
   {{\left(  2x_3 + 2x_2x_3 -1\right) }^2}{{\partial_2}^2}
}}
$$
\eqalign{
{\cal L}_3\ &=\
7 + 48x_1\partial_1 + 144{{x_1}^2}{{\partial_1}^2} +
  84\left( 1 + x_2 \right) \partial_2 -
  144x_1\left( 1 + x_2 \right) \partial_1\partial_2\cr& +
  36{{\left( 1 + x_2 \right) }^2}{{\partial_2}^2} +
  84x_3\partial_3 - 144x_1x_3\partial_1\partial_3 +
  72\left(  x_3 + x_2x_3 -1\right) \partial_2\partial_3 \cr&+
  36{{x_3}^2}{{\partial_3}^2}\ .
}$$
For $x_3\to 0$, these operators turn precisely into the PF operators
of the rigid SU(3) theory~! Recall \KLT\ that these are given by an
Appell system of type $F_4(a,b;c,c';\a,\b)$ of the form
\eqn\tilL{\eqalign{
\tilde{\cal L}_1&=\ta(\ta+c-1)-\a(\ta+\tb+a)(\ta+\tb+b)\cr
\tilde{\cal L}_2&=\tb(\tb+c'-1)-\b(\ta+\tb+a)(\ta+\tb+b)\ }}
(where $\ta\equiv\a\partial_\a$ etc.), with
$(a,b;c,c';\a,\b)=(-1/6,-1/6;2/3,1/2;1/x_1,{x_2}^2/x_1)$. It
immediately follows that the period vector has for
$\alpha'\rightarrow 0$ the rigid $SU(3)$ periods $a_i, a_{D,i}$,
$i=1,2$ among its components, and thus that the exact string theory
prepotential indeed contains the non-perturbative prepotential \KLT\
of the rigid $SU(3)$ theory. Note that the form of the variable $x_3$
reflects that, analogous to model $A$, string corrections break the
global $Z_{12}$ $R$-symmetry of the rigid Yang-Mills theory down to
$Z_6$.

Though we do not want to go into the details, we just note that we
find for the PF operators ${\cal L}_i$ four series and four
logarithmic solutions, each with indices $(-1/6,0,-1/3)$,
$(-1/6,0,1/3)$, $(-1/6,0,0)$ and $(-1/6,1,0)$. The series and
logarithmic solutions with the last two indices contain the rigid
$SU(3)$ periods. Specifically, after undoing the rescaling by
${z_1}^{1/6}\sqrt z_3$, we find a behavior of the PF
solutions that is fully compatible with a period vector of the
form $(1,S,\sqrt{\alpha'}a_1,\sqrt{\alpha'}a_2,
\sqrt{\alpha'}a_{D,1},\sqrt{\alpha'}a_{D,2},\alpha'u, \alpha' u S)$;
similar to the previous discussion, $u$ is a period due to the
fact\foot{This can be shown for all $SU(n)$ \ref\stef{S.\ Theisen,
private communication.}.} that $u=-2\pi i\partial_{S}F_{SW}^{SU(3)}$.

There are scaling arguments for model B that are similar to the one
for model $A$, with however a somewhat surprising conclusion.
That is, near the $SU(3)$ singular point the moduli scale as:
%
\eqn\psii{\eqalign{
\psi_0\ &=\ -2^{1/6}\epsilon^{1/12}\tu^{1/4}\cr
\psi_1\ &=\
-{{1\over\sqrt\epsilon}}-
\sqrt{\epsilon}\Big(\tu^{3/2}+
{3\over2}\sqrt3{\tilde v}\Big) \cr \psi_2\ &=\ -{{1}\over \epsilon}}}
Blowing up the singular neighborhood requires rescaling $z_3\to
\epsilon^{-1/12}z_3$, and after further irrelevant numerical
rescalings and setting $\zeta=z_1z_2$, we arrive at the following
representation of \MB\ near the singularity:
\eqn\popone{
\eqalign{
p_0\ &=\ {1\over 12}
\big(\zeta^{12}+ 2 \zeta^6 \zt3^6 + \zt3^{12}\big)\ =\ \epsilon\cr
p_1\ &=\ {1\over24}{\zt1}^{24}+ {1\over24}{\zt2}^{24}+
{1\over3}{\zt4}^{3}+ {1\over2}{\zt5}^{2} \cr&\qquad -
\big({1\over6}\tu^{3/2}+{\sqrt3\over4}{\tilde v}\big)
{\zt1}^{12}{\zt2}^{12}-
2^{1/6}i \tu^{1/4}{\zt1}^{2}{\zt2}^{2}{\zt4}{\zt5}=-1\ ,}}
where we have already eliminated $z_3$ from $p_1$ via solving
$p_0=0$. Like for model A, $p_0=0$ is a singular, rigid manifold, and
the sub-leading piece, $p_1=-1$, carries all moduli dependence.
However, the similarity stops there, in that $p_0=0$ does not
correspond to a singular $K3$, but to a singular point. Furthermore,
$p_1=-1$ can be compactified to a Calabi-Yau manifold by adding an
extra weight $2$ variable $w$, i.e. by considering $p_1+w^{12}=0$ in
the same weighted projective space as the original Calabi-Yau. This
``rigid'' Calabi-Yau manifold has no obvious relationship to the
hyperelliptic, genus two curve \refs{\KLTY{,}\AF}\ in
$W\!P^6_{1,1,3}$ that underlies the $SU(3)$ gauge theory!

Indeed, expecting to see the genus two curve is perhaps too much to
ask for\foot{Even for model A one can obtain, instead of the
Seiberg-Witten torus, a ``rigid'' Calabi-Yau in the
original weighted projective space that gives the same
periods with an appropriate meromorphic three-form.}. All that is
required is that the Calabi-Yau $p_1+w^{12}=0$ reproduces the $SU(3)$
quantum discriminant (which it does), and, more specifically, that it
has $(a_i,a_{D,i})$ among its periods. In other words, the
sub-leading piece in the degeneration of \MB\ is required to be
equivalent to the hyperelliptic curve only as far as the variation of
Hodge structure is concerned, since this is the only attribute that
is directly relevant for physical computations. Following reasoning
similar to that which led us to the meromorphic one-form for model A,
we expect that periods of a meromorphic three-form (obtained by
multiplying the holomorphic three-form of $p_1+w^{12}=0$ by
$w^{5}/(z_1^5z_2^5)$) provide an alternative description of $SU(3)$
quantum $N=2$ Yang-Mills theory.

We can get additional insight in the geometry of the specialization,
by studying the location of the nodes on the Calabi--Yau manifold in
relation to the other two types of semi-classical gauge symmetry
enhancement. These are characterized by the $SU(2)$ line: $T=U
\longleftrightarrow (\Delta _z=0,\ \Delta _x\neq0)$, and by the
$SU(2)\times SU(2)$ point: $T=U=i \longleftrightarrow
(\Delta _z=0,\ \Delta _x=0,\ x=2)$. A rough sketch of the
relevant part of the moduli space is given in Fig.2. Specifically,
the
coordinates of the nodes on the Calabi--Yau are
\eqn\nodes{\eqalign{
a)\ \Delta _z&=0:\ (1,1,\psi_1^{1/6},0,0)\cr
b)\ \Delta _x&=0:\ (1,1,(\psi_0^6+\psi_1)^{1/6},\psi_0^2
(\psi_0^6+\psi_1)^{1/3},\psi_0^3(\psi_0^6+\psi_1)^{1/2})\ ,}}
up to equivalences induced by phase symmetries (we have
displayed only one of two branches).

\goodbreak\midinsert
\centerline{\epsfxsize 2.3truein\epsfbox{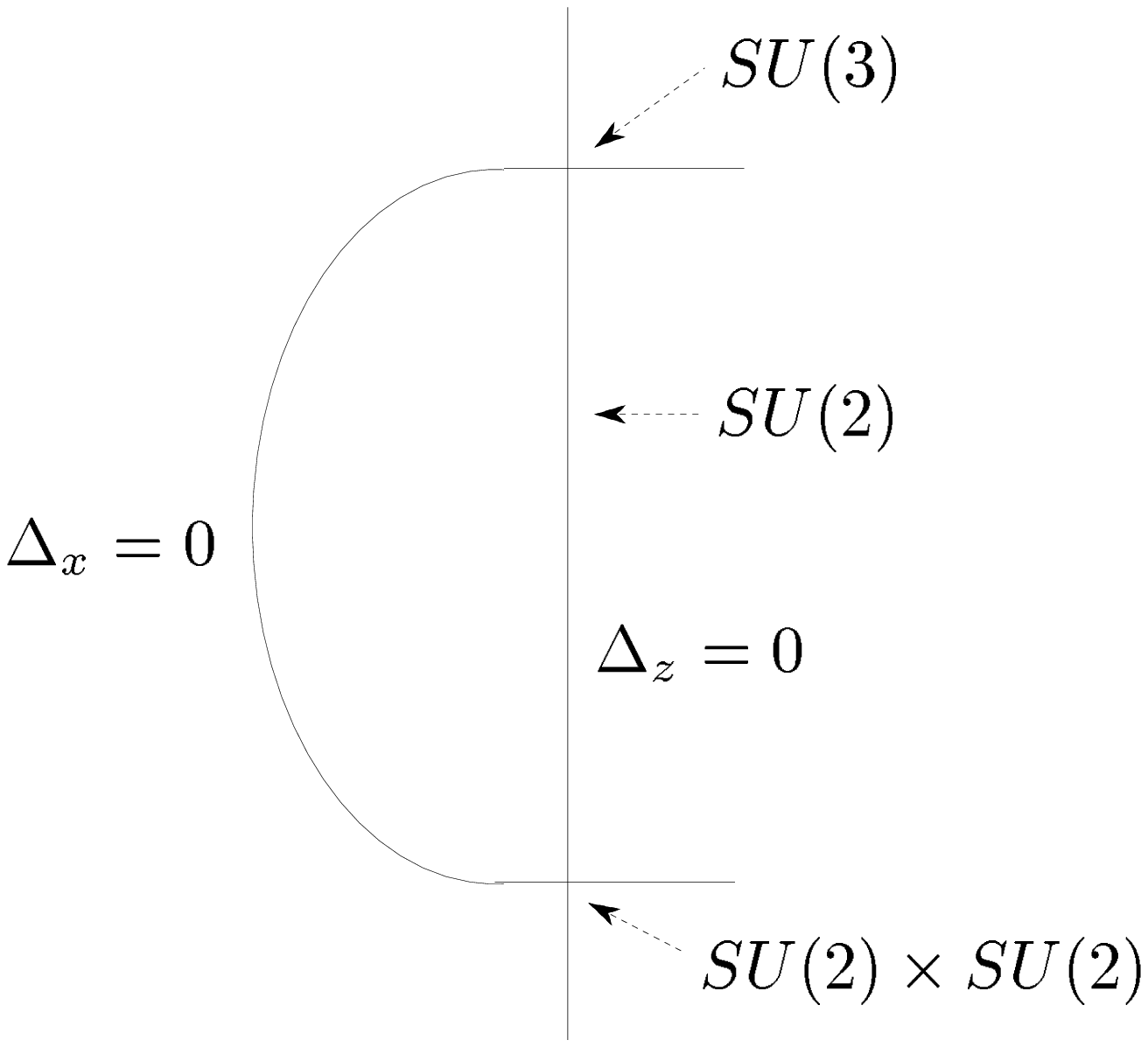}}
\smallskip
\leftskip 2pc \rightskip 2pc
\noindent{
\ninepoint\sl \baselineskip=8pt
{\bf Fig.2}
Shown is the singular locus for $y=0$, on which certain three-cycles
shrink to zero size and where various monopoles become massless.
We indicated the regions that are associated with semi-classical
gauge symmetry enhancements. Note that the line $\Delta_x=0$ is not
related to a semi-classical gauge symmetry, but rather is a pure
quantum effect.}
\smallskip\endinsert

For the $SU(2)\times SU(2)$ point, the relation between heterotic and
Calabi--Yau moduli can, in leading order, be inferred from:
$
j(T)=1728+432 \epsilon a_T^2,\ j(U)=1728+432 \epsilon a_U^2
$,
with $y=\epsilon^2$ and $a_T=a_1+a_2$ and $a_U=a_1-a_2$. If we
proceed as before and rescale the coordinates in order to make the
location of the nodes independent of the limit $\epsilon\to0$, we
find:
\eqn\pipapo{
\eqalign{
p_0\ &=\ {1\over12}y^{12}+{1\over12}{z_3}^{12}
+{1\over3}{z_4}^{3}+{1\over2}{z_5}^{2}+{1\over6}y^6{z_3}^{6}
-(2\a)^{1/6}y\,z_3 z_4 z_5\cr
p_1\ &=\ {1\over24}{z_1}^{24}+{1\over24}{z_2}^{24}+
{u_2\over12}(z_1z_2z_3)^6- {1\over24}2^{1/6}\a^{-5/6}(u_1+u_2)
z_1z_2z_3z_4z_5\ ,}}
where we have  set $y=z_1z_2$ and where we have used the
semi-classical relationship $u_i=a_i^2$. Specifically, the parameter
$\alpha$
takes a value equal to one, and it is precisely for this value that
the $K3$ surface develops {\it two} singularities, corresponding to
the nodes $a)$ and $b)$ \nodes\ of the Calabi--Yau manifold.
Inserting the coordinates of the singularities into $p_1$ as before,
we get two branches associated with
$$
p_1(i) = {1\over 24}z_1^{24}+{1\over 24}z_2^{24}
+{1\over 12}u_i\ z_1^{12}z_2^{12}\ ,\qquad\ i=1,2\ .
$$
That is, we obtain two copies of a six-fold cover of the
Seiberg--Witten torus \gammatwo. Note that the nodes are located at
points that are separated by an infinite distance when we take the global
limit.  It is plausible that it is this infinite distance
between the nodes in the global limit that leads to the two decoupled
$SU(2)$ factors.

For a generic point on the $SU(2)$ line, where $\Delta_z=0$, node
$b)$ is not developed and indeed, following similar reasoning, we
find just a single copy of the $SU(2)$ curve. As we approach the
$SU(3)$ point, node $b)$ develops, but in this case without
the rescaling of $z_4,z_5$ that infinitely separated the nodes at the
$SU(2)\times SU(2)$ point. The $SU(3)$ periods will arise in part
from integration contours that link both nodes, and apparently it is
due to the finite separation of the two nodes that we do not find a
genus two curve in a simple way.  More generally,
adopting this point of view suggests that while the rank of the gauge
group is given by the total number of nodes, the rank of a simple
group factor equals the number of a given set of nodes that are not
infinitely separated as $\alpha'\rightarrow 0$.

Finally we note that it is easy to check that appropriate meromorphic
forms for the tori come out in a way that is quite similar to what we
have discussed in section~3.

\bigskip
\centerline{{\bf Acknowledgements}}

We would like to thank S.\ Hosono for participation at the initial
stages of this work. In addition we would like to thank P.~Candelas,
M.~ Douglas, S.~Ferrara, S.~Katz, D.~L\"ust, K.S.~Narain, T. Taylor,
S.~Theisen, S.~Yankielowicz and S.-T.~Yau for discussions. AK, WL and
PM thank the Harvard Physics Department for hospitality.

The work of SK is supported in part by the
Harvard Society of Fellows and the William F. Milton Fund of Harvard
University.  The work of CV is supported in part by NSF grant
PHY-92-18167.

 \goodbreak

\listrefs
\end